\documentclass[conference, 10pt]{IEEEtran}
\IEEEoverridecommandlockouts
\usepackage{cite}
\usepackage{amsmath,amssymb,amsfonts}
\usepackage{algpseudocode}
\usepackage{graphicx}
\usepackage{textcomp}
\usepackage{xcolor}
\usepackage{braket}
\usepackage{url}
\usepackage{hyperref}
\usepackage{booktabs} % For professional looking tables
\usepackage{multirow} % For multi-row cells

\hypersetup{
  colorlinks   = true, %Colours links instead of ugly boxes
  urlcolor     = blue, %Colour for external hyperlinks
  linkcolor    = blue, %Colour of internal links
  citecolor   = blue %Colour of citations
}

\renewcommand{\figureautorefname}{Figure~\negthinspace}

\renewcommand{\tableautorefname}{Table~\negthinspace}

\def\BibTeX{{\rm B\kern-.05em{\sc i\kern-.025em b}\kern-.08em
    T\kern-.1667em\lower.7ex\hbox{E}\kern-.125emX}}
\begin{document}

\title{Programming Variational Quantum Circuits with Quantum-Train Agent \thanks{The views expressed in this article are those of the authors and do not represent the views of Wells Fargo. This article is for informational purposes only. Nothing contained in this article should be construed as investment advice. Wells Fargo makes no express or implied warranties and expressly disclaims all legal, tax, and accounting implications related to this article.}
}

\author{
\IEEEauthorblockN{
     Chen-Yu Liu \IEEEauthorrefmark{1}\IEEEauthorrefmark{6}\IEEEauthorrefmark{9}, 
    Samuel Yen-Chi Chen\IEEEauthorrefmark{3},
    Kuan-Cheng Chen\IEEEauthorrefmark{4}\IEEEauthorrefmark{5}\IEEEauthorrefmark{8},
    Wei-Jia Huang\IEEEauthorrefmark{9}, 
    Yen-Jui Chang \IEEEauthorrefmark{10}\IEEEauthorrefmark{11}
}

\IEEEauthorblockA{\IEEEauthorrefmark{1}Graduate Institute of Applied Physics, National Taiwan University, Taipei, Taiwan}
\IEEEauthorblockA{\IEEEauthorrefmark{9}Hon Hai (Foxconn) Research Institute, Taipei, Taiwan}
\IEEEauthorblockA{\IEEEauthorrefmark{3}Wells Fargo, New York, NY, USA}
\IEEEauthorblockA{\IEEEauthorrefmark{4}Department of Electrical and Electronic Engineering, Imperial College London, London, UK}
\IEEEauthorblockA{\IEEEauthorrefmark{5}Centre for Quantum Engineering, Science and Technology (QuEST), Imperial College London, London, UK}

\IEEEauthorblockA{\IEEEauthorrefmark{10}Quantum Information Center, Chung Yuan Christian University, Taoyuan City, Taiwan}
\IEEEauthorblockA{\IEEEauthorrefmark{11}Master Program in Intelligent Computing and Big Data, Chung Yuan Christian University, Taoyuan City, Taiwan}
\IEEEauthorblockA{Email:\IEEEauthorrefmark{6} d10245003@g.ntu.edu.tw, \IEEEauthorrefmark{8}kuan-cheng.chen17@imperial.ac.uk}

}

\maketitle
\begin{abstract}
 In this study, the Quantum-Train Quantum Fast Weight Programmer (QT-QFWP) framework is proposed, which facilitates the efficient and scalable programming of variational quantum circuits (VQCs) by leveraging quantum-driven parameter updates for the classical slow programmer that controls the fast programmer VQC model. This approach offers a significant advantage over conventional hybrid quantum-classical models by optimizing both quantum and classical parameter management. The framework has been benchmarked across several time-series prediction tasks, including Damped Simple Harmonic Motion (SHM), NARMA5, and Simulated Gravitational Waves (GW), demonstrating its ability to reduce parameters by roughly 70-90\% compared to Quantum Long Short-term Memory (QLSTM) and Quantum Fast Weight Programmer (QFWP) without compromising accuracy. The results show that QT-QFWP outperforms related models in both efficiency and predictive accuracy, providing a pathway toward more practical and cost-effective quantum machine learning applications. This innovation is particularly promising for near-term quantum systems, where limited qubit resources and gate fidelities pose significant constraints on model complexity. QT-QFWP enhances the feasibility of deploying VQCs in time-sensitive applications and broadens the scope of quantum computing in machine learning domains.

\end{abstract}


\section{Introduction}
%\KC{KC Rewrite}

Quantum Computing (QC) and Quantum Machine Learning (QML) have emerged as transformative fields with the potential to revolutionize computational methods across various domains \cite{biamonte2017quantum, dunjko2016quantum}. By leveraging quantum phenomena such as superposition and entanglement, QML algorithms can process information across multiple states simultaneously, offering a level of parallelism unattainable with classical systems \cite{lau2017quantum}. This capability has opened new avenues for applications like classification tasks\cite{mitarai2018quantum,chen2021end,chen2022quantumCNN,qmlapp2,chen2024compressedmediq}, reinforcement learning\cite{chen2020variational,chen2022variational,yun2023quantum,chen2024efficient}, time-series forecasting\cite{chen2022quantumLSTM,chen2022reservoir,lin2024quantum}, and notably, the forecasting and fitting of gravitational waves—a critical aspect in high-energy physics and cosmology\cite{di2024quantum,chen2024qcq}.

However, conventional QML approaches face significant challenges, particularly in encoding large datasets into quantum circuits. Techniques such as gate-angle encoding and amplitude encoding\cite{huang2021power} are constrained by the limited number of qubits and the short coherence times of current quantum hardware, hindering scalability and practical applicability to real-world, large-scale data problems. Moreover, the inference phase of trained QML models often requires access to quantum hardware—typically through cloud-based platforms—where hybrid quantum-classical computations are performed layer-by-layer. This dependency introduces inefficiencies in time-sensitive applications, such as real-time decision-making in autonomous systems.

To address these limitations, the Quantum-Train (QT) framework has been proposed as an innovative hybrid quantum-classical architecture \cite{liu2024training, liu2024quantum, liu2024qtrl, lin2024quantum, liu2024federated, liu2024quantum2, lin2024quantum2, liu2024quantum3}. The core concept of QT is to decouple quantum processing from direct data handling by employing a quantum neural network (QNN) to generate the weights of a classical machine learning model. This strategy effectively bypasses the data encoding bottleneck, as data is processed entirely within the classical model, eliminating the need for quantum data input. Additionally, it removes reliance on quantum hardware during the inference stage, resulting in a fully classical model post-training, which is particularly practical and efficient for near-term QML applications.

Building upon the QT framework, this paper introduces the Quantum-Train Quantum Fast Weight Programmers (QT-QFWP) as a novel solution for temporal and sequential learning tasks. The QT-QFWP utilizes a classical neural network—referred to as the ``slow programmer''—to dynamically adjust the parameters of a variational quantum circuit termed the ``fast programmer'', with the slow programmer being tuned using the QT technique. Instead of overwriting the fast programmer's parameters at each time step, the slow programmer generates incremental updates, allowing the fast programmer to efficiently incorporate past observations. This approach addresses the limitations of quantum recurrent neural networks (QRNNs), which often suffer from prolonged training times due to the need for backpropagation through time and extensive quantum circuit evaluations.

In this work, we apply the QT-QFWP to the forecasting and fitting of Damped Simple Harmonic Motion (SHM), NARMA5 series and simulated gravitational wave signals, demonstrating a proof of concept for using QML to solve problems in high-energy physics and cosmology—referred to as Quantum for High Energy Physics (Q4HEP)\cite{di2024quantum}. Our results indicate that the QT-QFWP not only overcomes the hardware limitations of conventional QML methods but also provides a scalable and efficient framework for complex temporal tasks in scientific domains.

Our numerical simulations show that the proposed QT-QFWP model performs exceptionally well in time-series prediction tasks, delivering results comparable to or surpassing those of QRNN-based models, such as the Quantum Long Short-Term Memory (QLSTM) network. Additionally, the QT-QFWP achieves performance on par with, and occasionally exceeding, that of the original QLSTM and QFWP model. By reducing the computational overhead of backpropagation-through-time calculations and minimizing the number of tunable parameters during training, the QT-QFWP framework provides a practical and efficient way to harness quantum advantages for sequential learning within the limitations of current quantum technologies.

\section{Related Works}
\label{sec:related_works}

Sequential data modeling, including time-series prediction and natural language processing, constitutes a crucial application area within QML. Recent advances in QML have been inspired by successful techniques from classical ML, particularly those based on recurrent neural networks (RNNs). Quantum RNNs have shown potential in effectively modeling time-series data \cite{bausch2020recurrent,chen2022quantumLSTM, chen2022reservoir} and addressing a variety of natural language tasks \cite{di2022dawn,stein2023applying,li2023pqlm}. However, training quantum RNNs can be time-consuming, primarily due to the requirement of backpropagation-through-time (BPTT) \cite{chen2022quantumLSTM} and the possible need for deep quantum circuits \cite{bausch2020recurrent}.

An alternative approach to sequential modeling in QML, without relying on quantum RNNs, is the \emph{quantum fast weight programmer} (QFWP) framework \cite{chen2024learning}. This framework introduces two networks: the \emph{slow programmer}, implemented using a classical NN, and the \emph{fast programmer}, realized with a QNN. At each time step, the slow programmer generates updates for the fast programmer based on the input signal, allowing the fast programmer to be \emph{partially} reprogrammed to process different time-step inputs. The fast programmer retains information from previous time steps in its parameters, as it is not fully rewritten.

This work distinguishes itself from the original QFWP by utilizing a QNN to generate the weights (or programs) of the slow programmer, meaning that the classical NN weights in slow programmer is from another QNN. As demonstrated in subsequent sections, employing a QNN for weight generation can significantly reduce the number of trainable parameters while maintaining or even improving model performance compared to the original approach.

\section{Quantum Neural Networks}

\emph{Variational quantum circuits} (VQCs), also known as \emph{parameterized quantum circuits} (PQCs), represent a specialized class of quantum circuits characterized by trainable parameters. These circuits have become foundational in the construction of \emph{quantum neural networks} (QNNs) within today’s hybrid quantum-classical computing framework \cite{bharti2022noisy}. Research has shown that VQCs can deliver specific types of quantum advantages \cite{abbas2021power,caro2022generalization,du2020expressive}, highlighting their potential in advancing QML applications.

A typical VQC consists of three core components: an \emph{encoding circuit}, a \emph{variational circuit}, and the concluding \emph{measurements}.

\begin{figure}[htbp]
\vskip -0.1in
\begin{center}
\includegraphics[width=0.85\columnwidth]{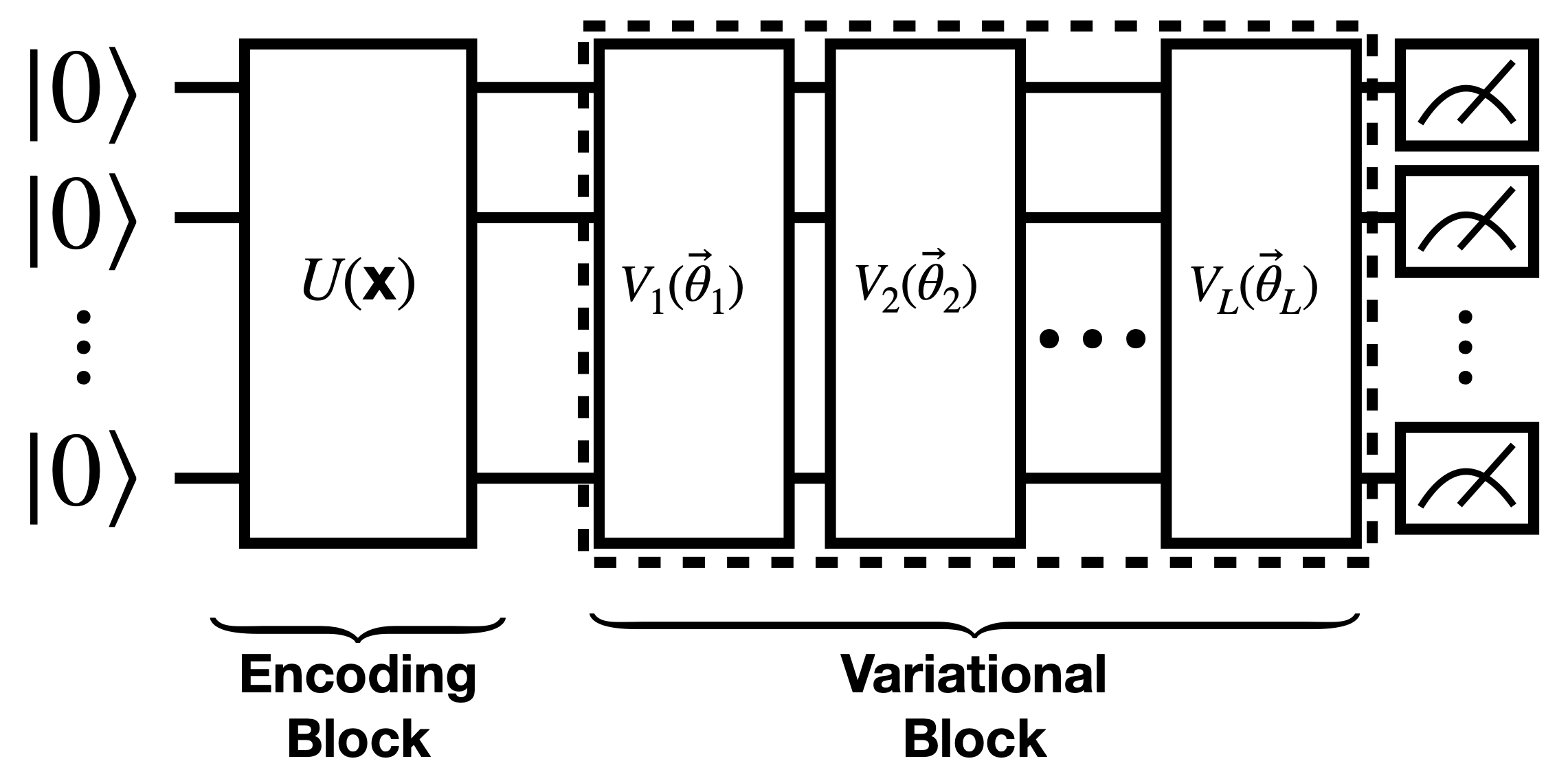}\vskip -0.1in
\caption{{\bfseries Generic Structure of a Variational Quantum Circuit (VQC).}}
\label{fig:generic_VQC}
\end{center}
\vskip -0.1in
\end{figure}

As illustrated in \figureautorefname{\ref{fig:generic_VQC}}, the encoding circuit $U(\mathbf{x})$ applies the input vector $\mathbf{x}$ to convert the initial quantum state $\ket{0}^{\otimes n}$ into the state $\ket{\Psi} = U(\mathbf{x})\ket{0}^{\otimes n}$. In this context, $n$ represents the number of qubits, and $U(\mathbf{x})$ denotes a unitary operator that is determined by the input $\mathbf{x}$.

\begin{figure}[htbp]
\vskip -0.1in
\begin{center}
\includegraphics[width=0.85\columnwidth]{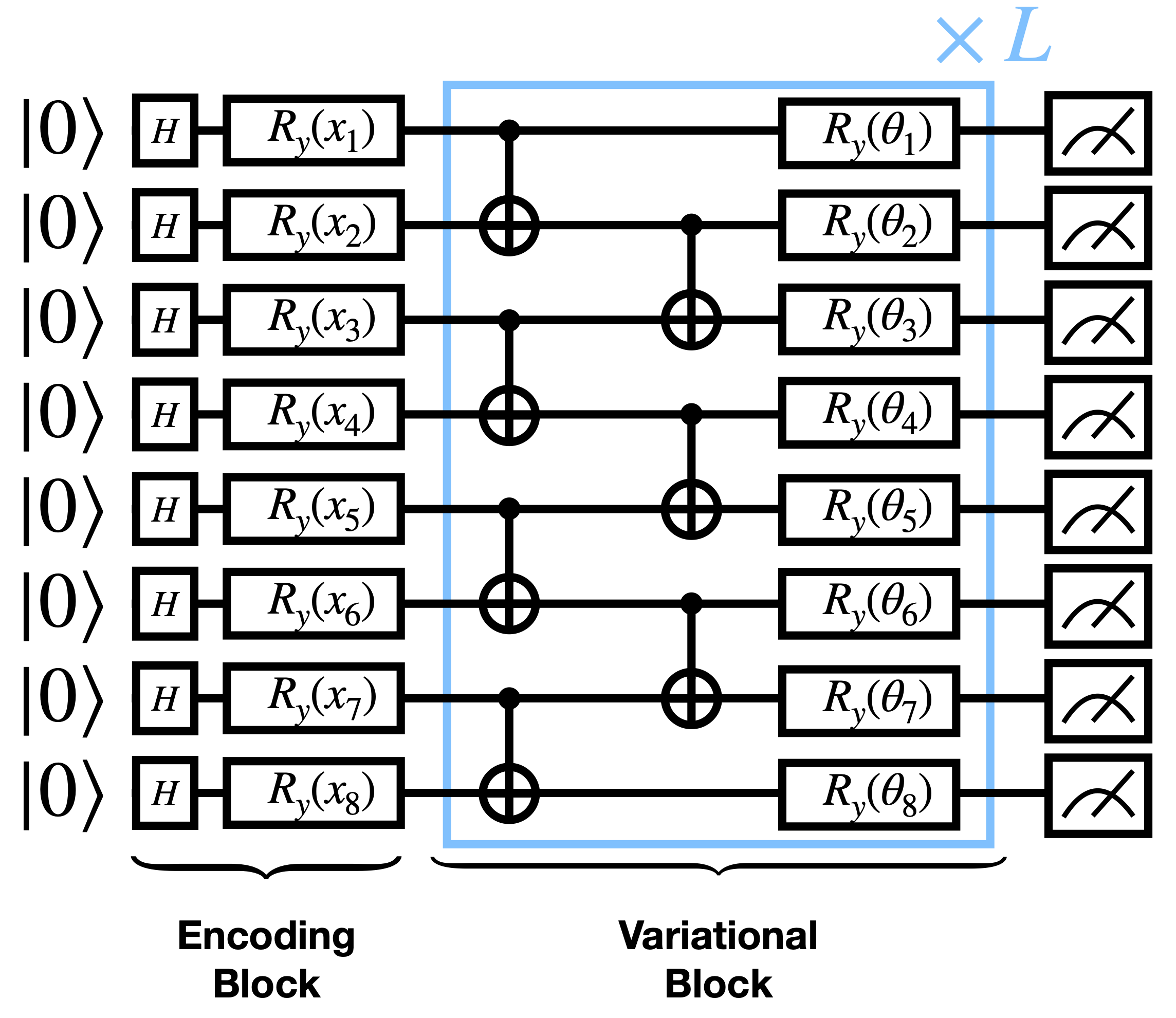}\vskip -0.1in
\caption{{\bfseries VQC used in this paper.}}
\label{fig:detailed_VQC}
\end{center}
\vskip -0.2in
\end{figure}

The VQC for the \emph{fast programmer} employed in this study is depicted in \figureautorefname{\ref{fig:detailed_VQC}}. The encoding block $U(\mathbf{x})$ consists of Hadamard gates applied to all qubits, initializing an unbiased state $H\ket{0}\otimes \cdots \otimes H\ket{0} = \sum_{(q_1, q_2, \dots, q_n) \in \{0,1\}^n} \frac{1}{\sqrt{2^n}} \ket{q_1} \otimes \ket{q_2} \otimes \cdots \otimes \ket{q_n}$. Additionally, $R_{y}$ gates are used, with rotation angles set according to the input data $x_1, \dots, x_n$, resulting in $U(\mathbf{x}) = R_{y}(x_1)H \otimes \cdots \otimes R_{y}(x_n)H$.

The encoded state then passes through the variational block (indicated by the dashed box), which consists of multiple layers of learnable quantum circuits $V_{j}(\vec{\theta_{j}})$. The $V_{j}$ circuit block used in this work is shown in the boxed region in \figureautorefname{\ref{fig:detailed_VQC}} and can be repeated $L$ times to increase the number of learnable parameters. Each $V_{j}$ block contains CNOT gates to entangle quantum information and parameterized gates $R_{y}$.

We denote the trainable part by $W(\Theta)$, defined as $W(\Theta) = V_{L}(\Vec{\theta_{L}})V_{L-1}(\Vec{\theta_{L-1}}) \cdots V_{1}(\Vec{\theta_{1}})$, where $L$ represents the number of layers and $\Theta$ is the collection of all trainable parameters $\Vec{\theta_{1}}, \cdots, \Vec{\theta_{L}}$. Each $\Vec{\theta_{k}}$ contains the parameters $(\theta_{1}, \cdots, \theta_{n})$ for that specific layer.

The VQC for the \emph{Quantum-Train} module, used to generate the neural network weights for the slow programmer, follows the same structure with Hadamard gates for system initialization and the variational blocks described earlier. The only distinction in the \emph{Quantum-Train} module is the absence of $R_y$ gates for data encoding.

In the \emph{fast programmer} VQC, the measurement process involves Pauli-$Z$ expectation values, with the output expressed as $\overrightarrow{f(\mathbf{x} ; \Theta)}=\left(\left\langle\hat{Z}_1\right\rangle, \cdots,\left\langle\hat{Z}_n\right\rangle\right)$, where $\left\langle\hat{Z}_{k}\right\rangle =\left\langle 0\left|U^{\dagger}(\mathbf{x})W^{\dagger}(\Theta) \hat{Z}_{k} W(\Theta)U(\mathbf{x})\right| 0\right\rangle$. Conversely, in the \emph{Quantum-Train} module, the probabilities of all $2^{n}$ computational basis states ($\ket{00 \cdots 0}, \cdots, \ket{11 \cdots 1}$) are obtained for subsequent use.

\section{Slow Programmer and Fast Programmer}
\label{sec:sp_fp}

\begin{figure}[htbp]
%\vskip -0.2in
\begin{center}
\includegraphics[width=0.85\columnwidth]{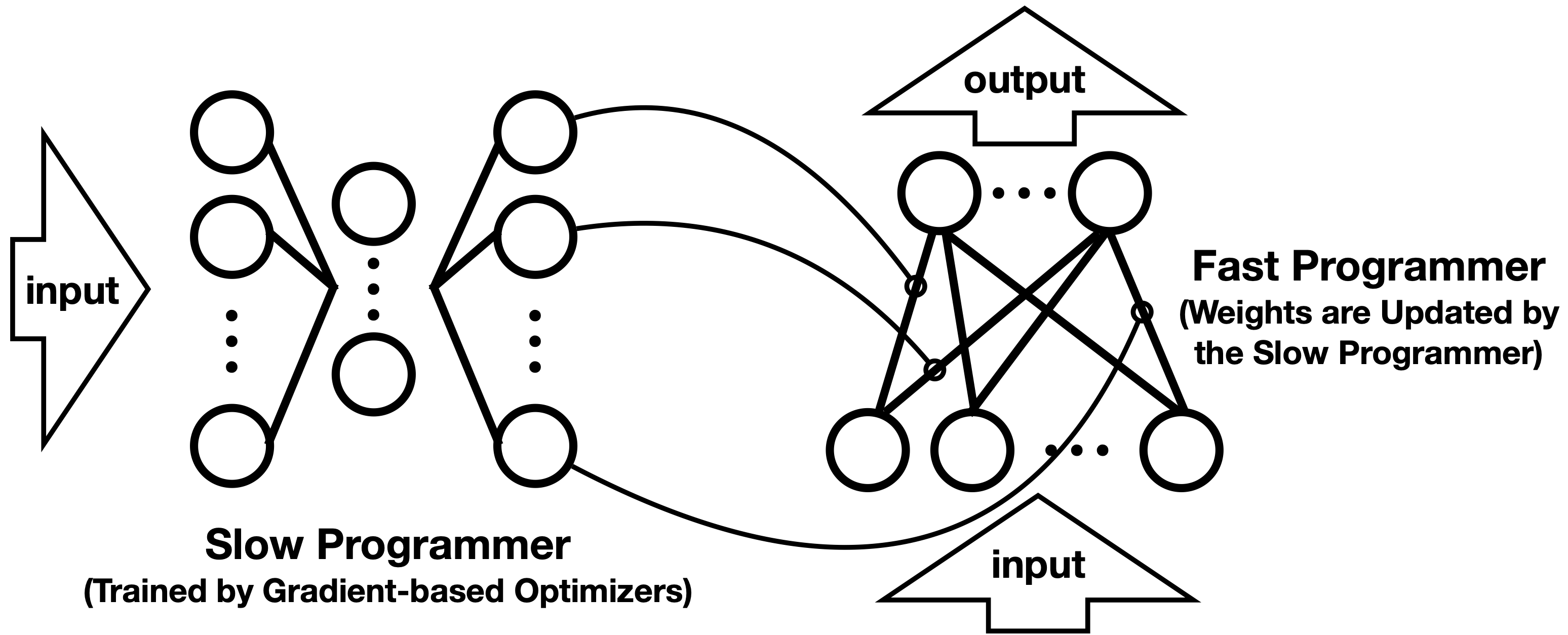}\vskip -0.1in
\caption{{\bfseries Generic Structure of a Fast Weight Programmer (FWP).}}
\label{fig:generic_FWP}
\end{center}
\vskip -0.2in
\end{figure}

The concept of \emph{Fast Weight Programmers} (FWP), depicted in \figureautorefname{\ref{fig:generic_FWP}}, was originally introduced in works by Schmidhuber \cite{schmidhuber1992learning,schmidhuber1993reducing}. In this sequential learning model, two distinct neural networks are used, termed the \emph{slow programmer} and the \emph{fast programmer}. Here, the neural network weights serve as the \emph{program} for the model or agent. The core idea of FWP involves the slow programmer generating \emph{updates} or \emph{modifications} to the fast programmer's weights at each time step, based on observations or inputs to the system. This \emph{reprogramming} process quickly refocuses the fast programmer on relevant information within the incoming data stream. Importantly, the slow programmer does not fully overwrite the fast programmer's weights; instead, only changes or updates are applied. This approach enables the fast programmer to incorporate prior observations or inputs (since it is not completely overwritten), allowing a simple feed-forward network to handle sequential prediction or control tasks without requiring RNNs, which often demand substantial computational resources.

In the original FWP configuration, each connection in the fast programmer is paired with a specific output unit from the slow programmer. The update rule governing the evolution of the fast programmer's weights is given by $W^{\text{fast}}(t+1) \leftarrow W^{\text{fast}}(t) + \Delta W(t)$, where $\Delta W(t)$ denotes the output of the slow programmer at time-step $t$.

The original scheme may face scalability challenges when the fast programmer NN is large, as the slow programmer requires an equal number of output neurons to match the connections in the fast programmer NN. An alternative approach, proposed in \cite{schmidhuber1992learning}, introduces a configuration where the slow NN includes dedicated units for each unit in the fast NN, labeled as \emph{FROM} and \emph{TO}. The \emph{FROM} units correspond to units from which at least one connection in the fast NN originates, while the \emph{TO} units correspond to those to which at least one fast NN connection leads. In this setup, weight updates for the fast NN are computed as $\Delta W_{ij}(t) = W^{\text{FROM}}_{i}(t) \times W^{\text{TO}}_{j}(t)$, where $\Delta W_{ij}(t)$ denotes the update for the fast NN weight $W^{\text{fast}}_{ij}(t)$. The entire FWP model can be optimized end-to-end through gradient-based or gradient-free methods. FWP has proven effective in addressing time-series modeling \cite{schmidhuber1992learning} and reinforcement learning tasks \cite{gomez2005evolving}.

\subsection{Quantum FWP}

In the quantum version of the FWP, the slow programmer—a classical NN—is configured to generate parameter updates for the fast programmer, which is implemented using a VQC \cite{chen2024learning}.

\begin{figure}[htbp]
\begin{center}
\includegraphics[width=1\columnwidth]{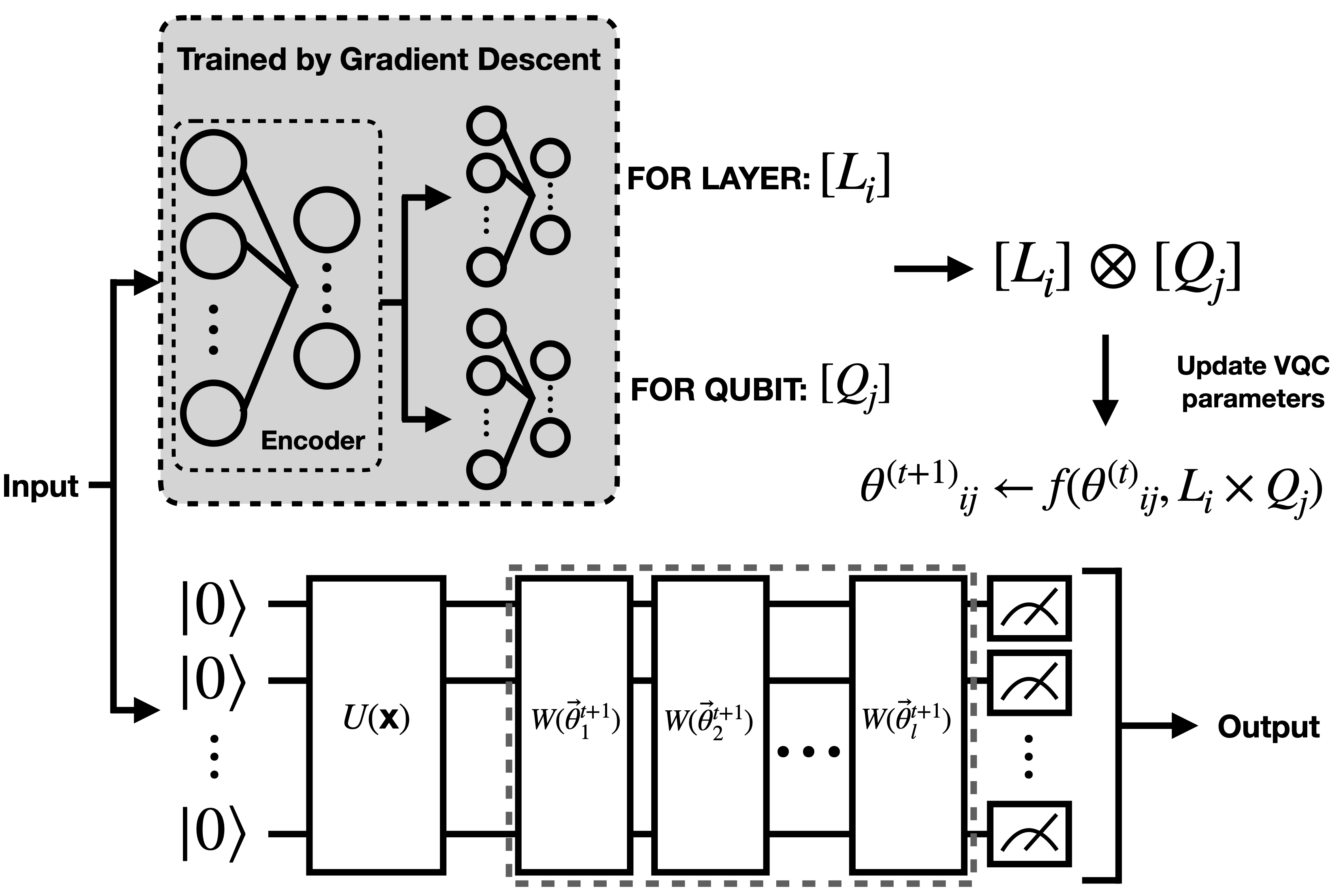}\vskip -0.1in
\caption{{\bfseries Quantum Fast Weight Programmers}}
\label{fig:vqFWP_Concept}
\end{center}
\vskip -0.2in
\end{figure}

As illustrated in \figureautorefname{\ref{fig:vqFWP_Concept}}, the input vector $\Vec{x}$ is initially processed by a classical neural network encoder. The output from this encoder is then passed through two additional neural networks. One network generates an output vector $[L_{i}]$ with a length corresponding to the number of VQC learnable layers, while the other network produces an output vector $[Q_{j}]$ with a length equal to the number of qubits in the VQC. The outer product of $[L_{i}]$ and $[Q_{j}]$ is then computed.
It can be expressed as $[L_{i}] \otimes [Q_{j}] = [M_{ij}] = [L_{i} \times Q_{j}] = 
\begin{bmatrix}
L_{1} \times Q_{1} & L_{1} \times Q_{2} & \cdots & L_{1} \times Q_{n}\\
L_{2} \times Q_{1} & L_{2} \times Q_{2} & \cdots & L_{2} \times Q_{n}\\
\vdots             &       \ddots       &        &        \vdots     \\
L_{l} \times Q_{1} & L_{l} \times Q_{2} & \cdots & L_{l} \times Q_{n}\\
\end{bmatrix}$, where $l$ is the number of VQC learnable layers in VQC and $n$ is the number of qubits.

At time $t+1$, the updated VQC parameters can be calculated as $\theta^{t+1}_{ij} = f(\theta^{t}_{ij}, L_{i} \times Q_{j})$, where $f$ combines the parameters from the previous time-step $\theta^{t}_{ij}$ with the newly computed term $L_{i} \times Q_{j}$. For the time-series modeling tasks considered in this work, we employ an \emph{additive} update rule, where the new circuit parameters are given by $\theta^{t+1}_{ij} = \theta^{t}_{ij} + L_{i} \times Q_{j}$. This approach allows information from previous time steps to be retained within the circuit parameters, influencing the VQC’s response when a new input $\Vec{x}$ is provided.

The output from the VQC can be further refined by additional components, such as scaling, translation, or a classical neural network, to enhance the final results.

\section{Quantum-Train Slow Programmer}
\label{sec:qt_sp}

While the QFWP proposal aims to generate the parameters of a VQC using a slow programmer classical model, QT, on the other hand, seeks to leverage the advantages of quantum computing to generate the parameters of a classical model through a combination of QNNs and classical mapping models \cite{liu2024quantum, liu2024training}.

\subsection{Quantum-Train}
For a target classical model with $p$ parameters, denoted as $\kappa \in \mathbb{R}^{p}$, QT employs a QNN with $n_{qt} = \lceil \log_2 p \rceil$ qubits. This QNN forms a Hilbert space of size $2^{\lceil \log_2 p \rceil} \geq p$, enabling it to produce $2^{n_{qt}}$ distinct measurement probabilities, expressed as  $|\langle \phi_i | \psi (\gamma) \rangle|^2 \in [0,1]$  for various basis states  $|\phi_i \rangle, \quad i \in \{1,2,...,2^{n_{qt}} \}$. These measurement probabilities are then mapped from $[0,1]$ to $\mathbb{R}$ through a mapping model $\mathcal{M}_{\beta}$ parameterized by $\beta$. This mapping model, implemented as a classical multi-layer perceptron (MLP), incorporates the information from the basis states  $|\phi_i \rangle$  to produce the corresponding parameter $\kappa_i$:
\begin{equation}
\mathcal{M}_{\beta}(|\phi_i \rangle, |\langle \phi_i | \psi (\gamma) \rangle|^2) = \kappa_i, \quad \forall i \in \{1,2,\ldots,p\}.
\end{equation}
As a result, the training process for the classical model parameters $\kappa$ involves optimizing both the QNN parameters $\gamma$ and the mapping model parameters $\beta$. Assuming that only $O(poly(n_{qt}))$ parameters \cite{cerezo2021variational, sim2019expressibility} are required for both  $|\psi(\gamma) \rangle$  and $\mathcal{M}_{\beta}$, the parameter reduction achieved by QT is expressed as  $p \rightarrow O(polylog(p)) $.

\subsection{Programming VQC with QT}

\begin{figure}[htbp]
\vskip -0.1in
\begin{center}
\includegraphics[width=1\columnwidth]{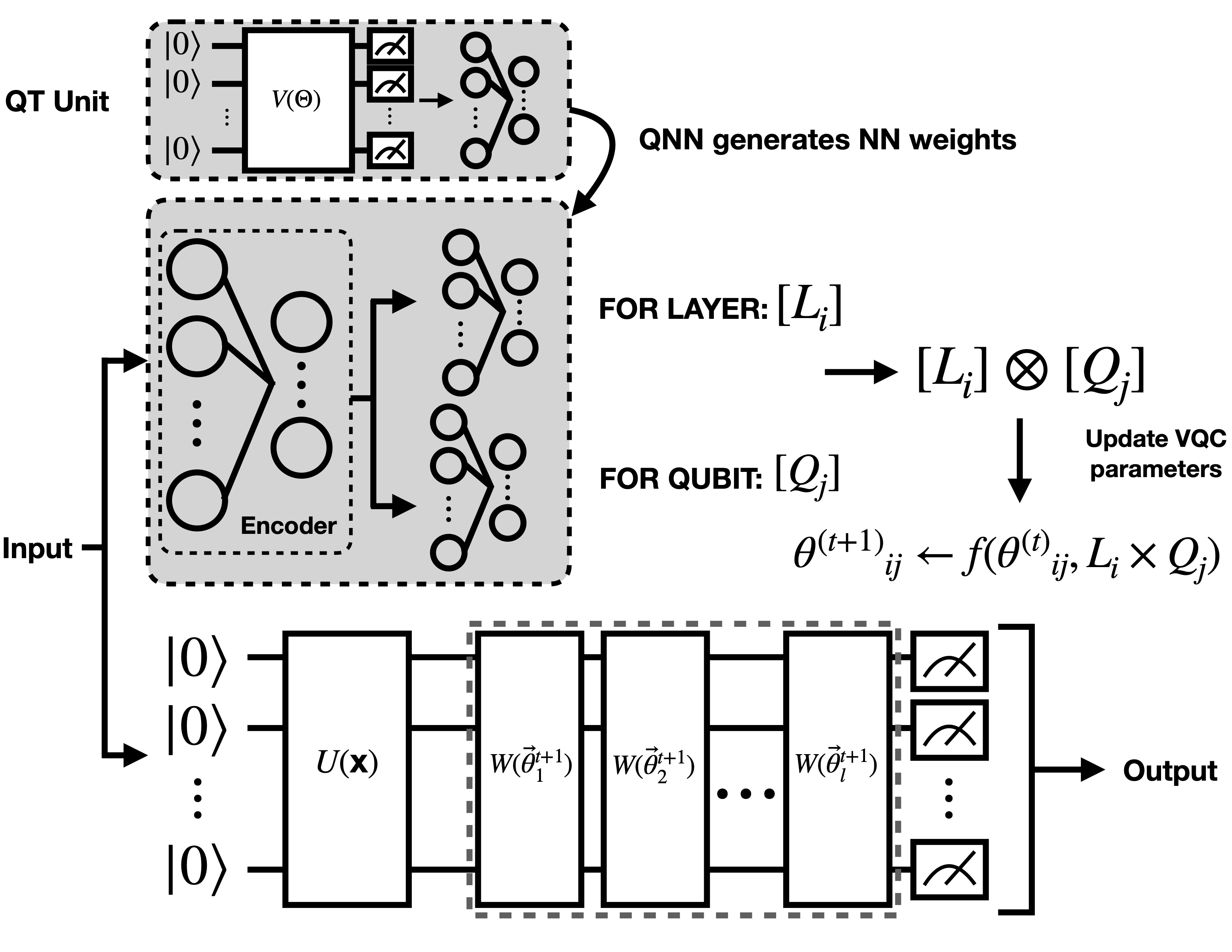}\vskip -0.1in
\caption{{\bfseries Quantum-Train Fast Weight Programmers (QT-QFWP)}}
\label{fig:qt_fwp_Concept}
\end{center}
\vskip -0.2in
\end{figure}

In QFWP, previous studies \cite{chen2024learning} utilize a classical slow programmer with 111 parameters to update a VQC with 16 parameters in time-series prediction tasks. In this setup, assuming the quality of the quantum computer is suitable for handling small QNN models (post-NISQ or early-fault-tolerant era), the number of training parameters for the slow programmer is significantly greater than that of the target VQC being updated. While this may raise concerns regarding the efficiency of such an approach, the overall number of training parameters is already reduced compared to the QLSTM model \cite{chen2022reservoir}, as demonstrated in the QFWP study. Our goal is to take this further—not only to reduce the training parameters compared to QLSTM but also to decrease the training parameters of QFWP. We aim to achieve this by applying the concept of QT, enabling the VQC to be programmed with QT, while, surprisingly, maintaining the model’s effectiveness. The idea of programming a VQC using QT combines the principles of QFWP and QT. In QFWP, the parameters of the VQC are updated by a classical model known as the slow programmer. We enhance this approach by \textit{employing QT to generate the parameters for the classical model}. In other words, the classical slow programmer is transformed into a \textit{Quantum-Train} slow programmer, as illustrated in \figureautorefname{\ref{fig:qt_fwp_Concept}}. Thus, we called this approach \textit{Quantum-Train Quantum Fast Weight Programmer} (QT-QFWP). 
This transformation leads to a further reduction in the number of training parameters on a polylogarithmic scale, as we will demostrate in the following sections.

\section{Numerical Results and Discussion}
\label{sec:nrd}

To demonstrate the applicability and effectiveness of the QT-QFWP algorithm in high-energy physics and cosmology, we have selected three datasets that encompass a range of complexities and physical phenomena. The Damped Simple Harmonic Oscillator represents fundamental physical systems, serving as a baseline for algorithm validation. The NARMA5 dataset introduces nonlinear and temporal dependencies, challenging the algorithm's capability to model complex, dynamic systems. Finally, the Gravitational Wave dataset simulates real-world astrophysical signals, providing a stringent test for the algorithm's proficiency in detecting and reconstructing signals of cosmological significance. Together, these datasets allow us to comprehensively assess the QT-QFWP algorithm's performance across different scenarios relevant to high-energy physics and cosmology.

\subsubsection{\textbf{Damped Simple Harmonic Motion}}
The Damped Simple Harmonic Motion (SHM) is a fundamental model in physics, describing systems where the restoring force is directly proportional to the displacement and acts in the opposite direction with external effect. The Damped SHM is governed by the second-order differential equation:
\begin{equation}
\frac{d^2x(t)}{dt^2} + 2 \zeta \omega \frac{d x}{d t} + \omega^2 x(t) = 0,
\end{equation}
where \( x(t) \) is the displacement at time \( t \), \( \omega = 2\pi f \) is the angular frequency of oscillation, and $\zeta = \frac{c}{2mk}$ is the damping ratio. 

For the purpose of evaluating the QT-QFWP algorithm, we generated a dataset comprising time-series data of Damped SHM systems as in the QFWP study\cite{chen2024learning}. Each data point consists of time \( t \) and the corresponding displacement \( x(t) \), sampled over multiple periods to capture the oscillatory behavior comprehensively. This dataset serves as a benchmark for assessing the algorithm's ability to accurately fit and predict harmonic motions.

\subsubsection{\textbf{NARMA5}}

The Nonlinear Autoregressive Moving Average model of order 5 (NARMA5) is a well-established benchmark for evaluating time-series prediction algorithms, especially in the context of nonlinear and dynamic systems. The NARMA5 sequence is defined recursively by:

\begin{align}
y(n+1) &= 0.3\, y(n) + 0.05\, y(n) \left( \sum_{i=0}^{4} y(n - i) \right) \nonumber \\
       &\quad + 1.5\, u(n - 4)\, u(n) + 0.1,
\end{align}

where:

\begin{itemize}
    \item \( y(n) \) is the system output at time step \( n \).
    \item \( u(n) \) is a sequence of independent random inputs uniformly distributed in the interval \([0, 0.5]\).
\end{itemize}

The complexity of NARMA5 stems from its dependence on both past outputs and inputs, making it an appropriate test case for evaluating the QT-QFWP algorithm's ability to handle nonlinear temporal dependencies.

The dataset was constructed by generating input sequences \( \{u(n)\} \) and computing the corresponding outputs \( \{y(n)\} \) using the recursive relation. This dataset captures the intricate dynamics of the NARMA5 system, providing a challenging scenario for curve fitting and prediction tasks.

\subsubsection{\textbf{Gravitational Wave Dataset}}

Gravitational Waves (GWs) are ripples in the fabric of spacetime produced by accelerating massive objects, such as merging black holes or neutron stars\cite{abbott2016observation}. Detecting GWs involves analyzing extremely weak signals buried in detector noise, posing significant challenges for data analysis algorithms.

The strain \( h(t) \) measured by a GW detector can be modeled as:

\begin{equation}
h(t) = F_+(\theta, \phi, \psi)\, h_+(t) + F_\times(\theta, \phi, \psi)\, h_\times(t) + n(t),
\end{equation}

where:

\begin{itemize}
    \item \( h_+(t) \) and \( h_\times(t) \) are the plus and cross polarization states of the GW signal.
    \item \( F_+ \) and \( F_\times \) are the detector's antenna response functions, dependent on the source's sky location \((\theta, \phi)\) and polarization angle \( \psi \).
    \item \( n(t) \) represents the detector noise.
\end{itemize}

The GW signals used in this study are simulated waveforms from binary black hole mergers, generated using inspiral-merger-ringdown models consistent with General Relativity, as described in the Kaggle page of the Riroriro GW simulation package\cite{van2021riroriro}. 

The objective is to detect and reconstruct the GW signals from the data using the QT-QFWP algorithm. Success in this task demonstrates the algorithm's potential contribution to the field of GW astronomy by enhancing signal detection and parameter estimation capabilities.

\subsection{Result Analysis}

\begin{table*}[!t]
\centering
\caption{Number of parameters in QLSTM, QFWP and QT-QFWP  models.}
\label{tab:parameters}
\begin{tabular}{|l|c|c|c|c|c|c|}
\hline
\multirow{2}{*}{} & \multicolumn{2}{c|}{QLSTM \cite{chen2022reservoir}} & \multicolumn{2}{c|}{QFWP \cite{chen2024learning}} & \multicolumn{2}{c|}{QT-QFWP}\\ \cline{2-7} 
                  & Classical & Quantum & Classical & Quantum & Classical & Quantum \\ \hline
Damped SHM & 5         & 144     & 111       & 16  & 14 & 23   \\ \hline
NARMA5    & 5         & 288     & 111       & 16  & 14 & 23    \\ \hline
Simulated GW    & --        & --     & --       & --  & 14 & 23    \\ \hline
\end{tabular}
\end{table*}

\begin{figure*}[htbp]
\vskip -0.1in
\begin{center}
\includegraphics[width=2\columnwidth]{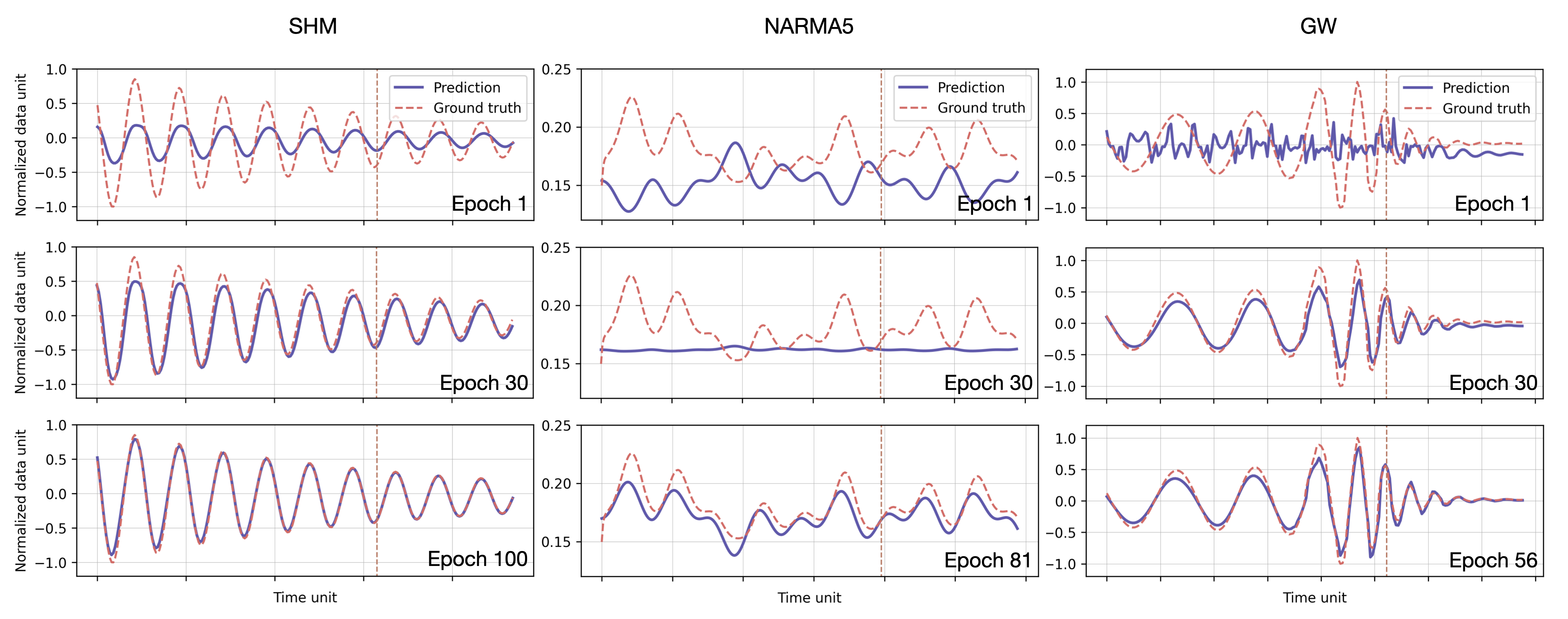}\vskip -0.1in
\caption{{\bfseries Results: QT-QFWP for damped SHM, NARMA5, and Simulated GW, fixing QT layers $=1$.}}
\label{fig:simu_res_1}
\end{center}
\vskip -0.2in
\end{figure*}

The proposed QT-QFWP approach was rigorously evaluated against the existing QFWP and QLSTM models across three distinct datasets: Damped Simple Harmonic Motion (SHM), NARMA5, and Simulated Gravitational Waves (GW). The primary objectives of this evaluation were to assess QT-QFWP's capability to achieve significant parameter compression without compromising predictive accuracy and to explore its potential in enhancing model describility, which are critical factors for applications in high-energy physics and cosmology.

Table~\ref{tab:parameters} demonstrates the parameter efficiency of each model. QT-QFWP consistently uses significantly fewer classical and quantum parameters compared to both QLSTM and QFWP across the Damped SHM and NARMA5 datasets. Specifically, QT-QFWP reduces the number of classical parameters by approximately 70–80\% compared to QFWP and reduces quantum parameters by roughly 85–90\% compared to QLSTM.
This substantial reduction is particularly beneficial in the context of quantum computing, where constraints on qubit resources and gate fidelities impose strict limits on model complexity. QT-QFWP generates the parameters for the classical slow programmer used in QFWP, where the slow programmer employs 106 parameters, as reported in the QFWP study \cite{chen2024learning}. Consequently, the required number of qubits for QT is $n_{qt} = \lceil \log_2 106 \rceil = 7$.
The classical parameter usage of QT-QFWP comes from the mapping model $\mathcal{M}_\beta$, which consists of $8 + 1 = 9$ parameters, and a post-processing neural network (with $4 \times 1 + 1 = 5$ parameters). The results presented in \tableautorefname{~\ref{tab:parameters}, \ref{tab:shm_results}, \ref{tab:narma5_results}, \ref{tab:gw_results}} and \figureautorefname{\ref{fig:simu_res_1}} use only a single layer of QNN in QT, comprising 7 parameters. Additionally, the target fast programmer is an 8-qubit VQC with 2 variational layers, contributing $8 \times 2 = 16$ parameters. Thus, the total number of quantum parameters is $16 + 7 = 23$.
The ability of QT-QFWP to achieve such compression without extensive parameterization underscores its efficiency and suitability for deployment on current and near-term quantum hardware.

\begin{table*}[!t]
\centering
\caption{Results: Time-Series Modeling - Damped SHM}
\label{tab:shm_results}
\begin{tabular}{|l|c|c|c|}
\hline
\multirow{2}{*}{} & QLSTM \cite{chen2022reservoir} & QFWP\cite{chen2024learning} & QT-QFWP   \\ \hline
Epoch 1           & $1.66 \times 10^{-1} / 1.35 \times 10^{-2}$ & $3.33 \times 10^{-1} / 3.26 \times 10^{-2}$ & $2.27 \times 10^{-1} / 1.98 \times 10^{-2}$  \\ \hline
Epoch 15          & $2.89 \times 10^{-2} / 5.53 \times 10^{-3}$ & $7.21 \times 10^{-2} / 1.65 \times 10^{-2}$ & $3.05 \times 10^{-2} / 5.78 \times 10^{-3}$ \\ \hline
Epoch 30          & $9.06 \times 10^{-3} / 3.41 \times 10^{-4}$ & $5.96 \times 10^{-2} / 1.34 \times 10^{-2}$ & $2.37 \times 10^{-2} / 5.15 \times 10^{-3}$\\ \hline
Epoch 100         & $2.86 \times 10^{-3} / 1.94 \times 10^{-3}$ & $1.09 \times 10^{-2} / 2.70 \times 10^{-3}$ & $1.57 \times 10^{-3} / 2.92 \times 10^{-5}$ \\ \hline
\end{tabular}
\end{table*}

\begin{table*}[!t]
\centering
\caption{Performance Comparison on NARMA5 Dataset}
\label{tab:narma5_results}
\begin{tabular}{|l|c|c|c|}
\hline
\multirow{2}{*}{} & QLSTM \cite{chen2022reservoir} & QFWP\cite{chen2024learning} & QT-QFWP   \\ \hline
Epoch 1           & $3.99 \times 10^{-3} / 4.07 \times 10^{-4}$ & $4.44 \times 10^{-2} / 3.48 \times 10^{-4}$ & $2.04 \times 10^{-2} / 1.20 \times 10^{-3}$ \\ \hline
Epoch 15          & $3.30 \times 10^{-4} / 4.23 \times 10^{-4}$ & $2.99 \times 10^{-4} / 8.76 \times 10^{-5}$ & $4.03 \times 10^{-4} / 7.33 \times 10^{-4}$\\ \hline
Epoch 30          & $1.86 \times 10^{-4} / 2.06 \times 10^{-4}$ & $2.71 \times 10^{-4} / 1.57 \times 10^{-4}$ & $3.40 \times 10^{-4} / 5.39 \times 10^{-4}$\\ \hline
Epoch 100         & $9.85 \times 10^{-5} / 2.52 \times 10^{-5}$ & $5.15 \times 10^{-5} / 1.68 \times 10^{-5}$ & $1.32 \times 10^{-4} / 1.98 \times 10^{-4}$\\ \hline
\end{tabular}
\end{table*}

\begin{figure*}[htbp]
\vskip -0.1in
\begin{center}
\includegraphics[width=2\columnwidth]{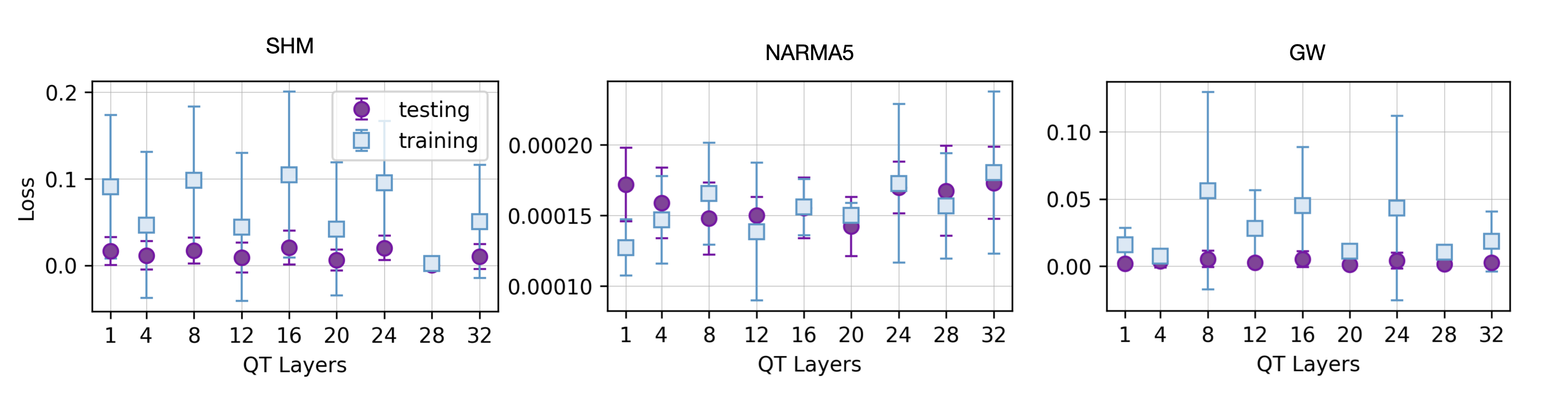}\vskip -0.1in
\caption{{\bfseries Results: Different QT layers of QT-QFWP for damped SHM, NARMA5, and Simulated GW.}}
\label{fig:simu_res_2}
\end{center}
\vskip -0.2in
\end{figure*}

In the context of the Damped SHM dataset, which serves as a foundational benchmark for evaluating models' capacity to capture linear dynamical systems, QT-QFWP demonstrates competitive performance metrics. As shown in Table~\ref{tab:shm_results} and \figureautorefname{\ref{fig:simu_res_1}}, QT-QFWP achieves a training loss of \(1.57 \times 10^{-3}\) and a testing loss of \(2.92 \times 10^{-5}\) at epoch 100, outperforming QLSTM, which records an training loss of \(2.86 \times 10^{-3}\) and an testing loss of \(1.94 \times 10^{-3}\). This indicates that QT-QFWP not only maintains but also enhances predictive accuracy despite a significant reduction in model parameters, suggesting an effective utilization of the quantum feature space to capture the underlying physical dynamics.

The evaluation on the NARMA5 dataset, characterized by its nonlinear and temporal dependencies, presents a more complex challenge for time-series modeling. According to Table~\ref{tab:narma5_results} and \figureautorefname{\ref{fig:simu_res_1}}, while QFWP achieves the lowest training and testing loss at epoch 100 (\(5.15 \times 10^{-5}\) and \(1.68 \times 10^{-5}\), respectively), QT-QFWP maintains competitive performance with an training loss of \(1.32 \times 10^{-4}\) and an testing loss of \(1.98 \times 10^{-4}\). Although QT-QFWP does not attain the absolute lowest error metrics in this scenario, the slight increase in error is considerably offset by the dramatic reduction in parameter count. This highlights QT-QFWP's robustness in managing complex, nonlinear systems efficiently, presenting a favorable trade-off between accuracy and model complexity, especially in environments constrained by quantum hardware limitations.

The Simulated GW dataset represents a real-world application involving the detection and reconstruction of gravitational wave signals. As detailed in Table~\ref{tab:gw_results} and \figureautorefname{\ref{fig:simu_res_1}}, QT-QFWP exhibits substantial improvements in both training and testing loss across training epochs, culminating in an training loss of \(6.44 \times 10^{-3}\) and an testing loss of \(1.52 \times 10^{-3}\) at epoch 100. These results demonstrate QT-QFWP's capability to accurately model and reconstruct complex astrophysical signals with a minimal parameter set, thereby affirming its potential utility in high-energy physics and cosmological studies where efficient data processing is paramount.

\begin{table}[!b]
\centering
\caption{Performance of QT-QFWP on Simulated GW Dataset}
\label{tab:gw_results}
\begin{tabular}{|l|c|}
\hline
\multirow{2}{*}{} & QT-QFWP   \\ \hline
Epoch 1           & $2.37 \times 10^{-1} / 4.58 \times 10^{-2}$\\ \hline
Epoch 15          & $8.51 \times 10^{-2} / 1.13 \times 10^{-2}$ \\ \hline
Epoch 30          & $3.59 \times 10^{-2} / 6.23 \times 10^{-3}$ \\ \hline
Epoch 100         & $6.44 \times 10^{-3} / 1.52 \times 10^{-3}$ \\ \hline
\end{tabular}
\end{table}

While above results utilize only single layer QNN in the QT part, it is also possible to extend the number of QNN layer in QT (expressed as QT layers), to explore the effect of deeper QNN on the performance, where the quantum parameter usage is scaled as $16 + 7 \times N_{\text{QT layers}}$. \figureautorefname{\ref{fig:simu_res_2}} illustrates the performance of QT-QFWP across different numbers of QT layers for three datasets: Damped SHM, NARMA5, and Simulated GW. The y-axis represents the loss, while the x-axis indicates the number of QT layers, with both training and testing losses visualized for each configuration. 

Across all three datasets, both training (blue squares) and testing (purple circles) losses generally remain stable as the number of QT layers increases. The Damped SHM dataset shows relatively low loss values overall, with minimal variation between training and testing, indicating that the model generalizes well. For the NARMA5 dataset, both training and testing losses are small, but the error bars reveal some variability, particularly for higher QT layers, suggesting potential overfitting risks as layer depth increases. In the Simulated GW dataset, the losses start very low and remain consistent even with an increasing number of QT layers, reflecting a high model effectiveness with minimal overfitting. 

These results also indicate that QT-QFWP performs well even with a single-layer QNN, which is noteworthy given the hybrid model’s size. This efficiency stems from the advanced parameter generation technique that leverages the expanded capacity of the Hilbert space. Such a design demonstrates the strength of QT-QFWP in maintaining performance with minimal computational overhead, making it well-suited for resource-constrained quantum environments.

Overall, the empirical results unequivocally demonstrate that QT-QFWP achieves significant parameter compression without compromising, and in certain instances enhancing, model performance. The ability to reduce both classical and quantum parameters by up to 85--90\% while maintaining low error metrics is particularly beneficial for quantum computing applications, where resource constraints are a critical consideration. Additionally, the enhanced describility of the QT-QFWP models facilitates better interpretability of the underlying physical phenomena, which is essential for scientific research and the validation of theoretical models.

From a scientific standpoint, the successful application of QT-QFWP to both fundamental physical systems and complex real-world datasets such as gravitational waves signifies its versatility and robustness. The algorithm's efficiency in parameter usage aligns with the practical constraints of current quantum hardware, enabling more sophisticated analyses in high-energy physics and cosmology. Furthermore, the ability to enhance model describility contributes to the development of interpretable quantum machine learning models, fostering greater trust and reliability in their applications within scientific research.

\section{Conclusion and Future Work}
\label{sec:cfw}

In conclusion, QT-QFWP presents a compelling advancement in quantum machine learning by effectively balancing model complexity with predictive performance. As summarized in Table~\ref{tab:parameters}, from QLSTM to QFWP, the study of QFWP reduce the usage of quantum parameters by introducing more classical parameters (slow programmer) to control these quantum parameters (fast programmer), from QFWP to QT-QFWP, we introduce QT to further reduce the classical parameter usage by utilizing another QNN to control the classical slow programmer, as shown in \figureautorefname{~\ref{fig:qt_fwp_Concept}}. Its demonstrated efficacy across varied datasets highlights its potential to make substantial contributions to the fields of quantum computing, high-energy physics, and cosmology, particularly in scenarios where efficient and accurate time-series modeling is essential. 

Future work will focus on exploring the impact of deeper QNN architectures on the framework’s performance with more complicated tasks, as indicated by the analysis of different QT layers. Additionally, further research will investigate the use of advanced optimization techniques to mitigate potential overfitting risks observed in more complex datasets, such as NARMA5. Applying QT-QFWP to new fields, particularly quantum chemistry, offers a promising direction. In quantum chemistry, where deep quantum circuit ansatz often result in challenging optimization problems, QT-QFWP shows potential to reduce the number of training parameters in these circuits while maintaining effectiveness. This improvement could alleviate optimization bottlenecks and expand the framework’s applicability to molecular simulations and other quantum chemical tasks. As quantum hardware continues to evolve with improved gate fidelities and longer coherence times, the framework will be adapted to leverage these advancements, ensuring that QT-QFWP remains relevant and effective. These efforts will not only enhance the practical utility of quantum machine learning but also foster deeper integration of hybrid quantum-classical architectures in emerging technological landscapes.

\bibliographystyle{IEEEtran}
\bibliography{bib/tools,bib/vqc,bib/qml_examples,bib/quantum_fl, bib/ml_examples, bib/hybrid_co_examples,bib/classical_fl,references,bib/fwp,bib/qt}

\end{document}